# The lower symmetry electron-density distribution and the charge transport anisotropy in cubic dodecaboride LuB$_{12}$

Nadezhda B. Bolotina [1,*], Alexander P. Dudka [1], Olga N. Khrykina [1], Vladimir N. Krasnorussky [2], Natalya Yu. Shitsevalova [3], Volodymyr B. Filipov [c], Nikolay E. Sluchanko [2,4]

[1] Shubnikov Institute of Crystallography of Federal Scientific Research Centre 'Crystallography and Photonics' of Russian Academy of Sciences, 59 Leninskii Ave., 119333 Moscow, Russia

[2] Prokhorov General Physics Institute, Russian Academy of Sciences, 38 Vavilov Str., 119991 Moscow, Russia

[3] Frantsevich Institute for Problems of Materials Science, National Academy of Sciences of Ukraine, 3 Krzhyzhanovsky Str., 03680 Kiev, Ukraine

[4] Moscow Institute of Physics and Technology (State University), 9 Institutskiy Per., 141700 Dolgoprudny, Russia

**Abstract**

High-quality single crystals of LuB$_{12}$ are grown using the induction zone melting method. The X-ray data are collected at temperatures 293, 135, 95, 50 K. The crystal structure of LuB$_{12}$ can be refined with record low R-factor in the cubic $Fm\bar{3}m$ symmetry group despite reiterated observations of the cubic symmetry distortions both in the unit-cell values and in the physical properties. A peculiar computing strategy is developed to resolve this contradiction. True symmetry of the electron-density distribution in LuB$_{12}$ is proved to be much lower than cubic as a result, which correlates very accurately with anisotropy of transport properties of LuB$_{12}$.

## 1. Introduction

Dodecaborides $R$B$_{12}$ ($R$ = Y, Zr, Tb – Lu) possess unique combination of charge-transport, magnetic, thermal and mechanical properties [1] and have a simple *fcc* structure, which is similar to NaCl when sodium and chlorine atoms are replaced by the metal atoms and centers of regular cuboctahedra B$_{12}$, respectively (Fig. 1a). The metal cations center octahedral cavities formed by closely packed anionic groups B$_{12}$. In other presentation, the boron atoms form large truncated octahedra B$_{24}$ centered by the metal atoms and separated by empty cuboctahedra B$_{12}$ (Fig. 1b).

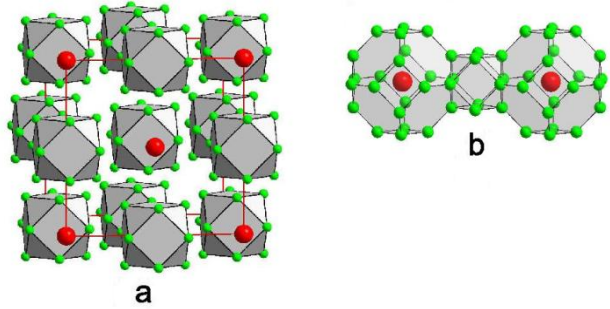

**Fig. 1**. (a) a NaCl-like unit cell of LuB$_{12}$, with Lu (red spheres online) as Na and B$_{12}$ clusters (B atoms shown as green spheres) as Cl. (b) two large B$_{24}$ polyhedra centered by Lu atoms and a smaller B$_{12}$ cuboctahedron between them.

Owing to the simple structure, dodecaborides $R$B$_{12}$ (LuB$_{12}$ in this number) are excellent models to study anomalies of various physical properties of the metal-like compounds [2–4]. Structural symmetry of $R$B$_{12}$ ($R$ = Y, Zr, Tb – Lu) is determined as cubic $Fm\bar{3}m$ [1]. ScB$_{12}$ falls out from this family, since its structure at normal conditions is determined by some researchers as cubic [5] but by other ones as tetragonal [6, 7]. Polymorphous phase transition from high-temperature *fcc* phase ($Fm\bar{3}m$) into low-temperature trigonal phase ($R\bar{3}m$) is observed in the Sc-containing solid solutions Zr$_{1-x}$Sc$_x$B$_{12}$, $0.1 \leq x \leq 0.9$. The threefold axis of the trigonal phase is oriented along a space diagonal of slightly distorted cube [8]. Morphotropic (dependent on composition at fixed temperature) phase transition from cubic to tetragonal phase is observed in several series of solid solutions $Me_{1-x}$Sc$_x$B$_{12}$, $0.1 \leq x \leq 0.95$ ($Me$ = Y, Zr, Tm, Lu) [9].

The cubic lattice of LuB$_{12}$ is most probably distorted. It concerns cationic sublattice above all, since large volumes of B$_{24}$ cavities are quite capable for both dynamic and static disorder of the Lu$^{3+}$ cations even on the assumption that the boron covalent framework is fixed. Raman spectra of ZrB$_{12}$ and LuB$_{12}$ contain modes, which would be forbidden by selection rules if the metal atoms occupied centers of inversion [10, 11]. Some disordering of electron density (ED) in the neighborhood of cationic sites was revealed in [10] using structure analysis. Similar distortions of local structure $R$B$_{12}$ ($R$ = Ho – Lu) were fixed by EXAFS spectroscopy [12]. Regular signals about probable symmetry distortions come from the researchers studying physical properties of dodecaborides. For instance, noticeable anisotropy of magnetoresistance has been revealed in LuB$_{12}$ [13]. As reported in [14], where the narrow-gap semiconductor YbB$_{12}$ was studied by the technique of electron paramagnetic resonance, the Yb$^{3+}$ – Yb$^{3+}$ pairs appeared at low temperatures along certain directions being accompanied by shifts of the cations from the inversion centers.

Other observations supply data on the unit-cell values of dodecaborides. Tetragonal distortions $a \cong b \neq c$ over the temperature range 90 – 300 K, more pronounced below 150 K, have been revealed in LuB$_{12}$. It is interesting that according to [1] the parameters yield to the formula $a \cong b < c$ whereas the study [15] shows that $a \cong b > c$ is more appropriate. Reliable estimate is a demanding task because of an extremely small difference between these parameters that does not exceed 0.002-0.003 Å in the temperature interval. Measurements are performed at different temperatures and the parameters can change with temperature in addition. It should nevertheless be noted that last ratio $a \cong b > c$ kept for several crystalline samples of LuB$_{12}$ measured on different-type diffractometers.

Cubic symmetry of LuB$_{12}$ is thus disputable. In spite of this, the single-crystal structure of LuB$_{12}$ has been refined not long ago in the $Fm\overline{3}m$ symmetry group with exceptionally low value of residual factor R = 0.2% [16]. This contradiction has been naturally resolved using original approach developed in this study. A lower symmetry ED distribution (charge stripes) in LuB$_{12}$ is revealed, which correlates very well with the filamentary structure of conduction channels observed in the magnetoresistance measurements.

## 2. Experimental

*2.1. Crystal growth, sample preparation and resistance measurements*

High-quality single crystals were grown using the induction zone melting method in an inert gas atmosphere from the preliminarily synthesized LuB$_{12}$ powders, see [17] for details. Evidently, reliable structure results call for reliable intensities of the diffraction reflections, which should be cleansed from the influence of all factors except the atomic and domain structure of the crystal. Thus, for example, the intensity divergence of equivalent reflections can be a consequence of the X-ray absorption anisotropy in a free-form crystal, so crystals were thoroughly treated to put them into almost ideal spheres. At first, they were cut into plates of 1 mm thickness in the electric-spark machine. Then, the plates were polished and further were cut into cubes with a side of approximately 0.85 mm. The cubes were rolled using the diamond and el'bor abrasive papers with grains from 80 to 3 microns to get spherical, barely ellipsoidal shapes and smooth surfaces whose roughness did not exceed 3 microns. Finally, the crystals were etched in the boiling solution of nitric acid $HNO_3:H_2O = 1:1$ to remove the distorted surface layer. Measurements of resistance and the transverse magnetoresistance were performed in a four-terminal scheme with a direct current **I** ⊥ **B**, **I** ∥ [001] axis in the crystal, at temperatures in the range 1.8 – 300 K and in a magnetic field up to B = 8 Tesla with the help of installation with a sample rotating in a magnetic field around the current axis [18].

*2.2. Data collection, data reduction and structure refinement*

Several X-ray data sets were collected from two single crystals of $LuB_{12}$ using three different-type diffractometers to assure the independence of the measurement results from the crystal sample or diffractometer. It should be noted that diffractometers with pointed detectors win on the measurement accuracy but lose on the data collection rate to diffractometers equipped with coordinate detectors. The data set from the crystal #1 was obtained at 293 K on a CAD4 diffractometer equipped with a pointed detector. Just this data set was used before to refine the $LuB_{12}$ structure in the $Fm\bar{3}m$ symmetry group with the record low value of the residual factor R = 0.2% [16]. The sample #1 was also used for the data collection at 50 K on a Huber 5042 diffractometer equipped with a pointed detector and a helium closed-cycle cryostat Displex DE-202. Although we studied before the structure of $LuB_{12}$ at 50 K [19], methods of the sample preparation and data collection on the Huber diffractometer were enriched with new details, so the new crystal #1 was measured using advanced techniques [15, 20, 21].

The crystal symmetry of $LuB_{12}$ at 50 K was last considered in [19] as tetragonal. This choice was motivated by strongly diverging intensities of the reflections, which were symmetry-equivalent in the cubic ($m\bar{3}m$) Laue class. As we learned by next experience measuring other $LuB_{12}$ samples at various temperatures, an internal factor $R_{int}$ of the reflection averaging was not a faultless indicator of the dodecaboride symmetry. Systematical errors (sources of some of them remain unknown) have an additional influence on $R_{int}$. This can be balanced to a variable degree when using data of different redundancy, which is higher in more symmetrical models. For instance, data collected from the crystal #1 at 50 K were averaged in the $m\bar{3}m$ Laue class with $R_{int}$ = 11.9% but then refined in $Fm\bar{3}m$ with R = 1.4% (see Table 1).

The X-ray data from the crystal #2 were collected at temperatures 135 K and 95 K using a coordinate diffractometer Xcalibur S [22]. General data on the crystals, experiments and the structure refinement results are summarized in Table 1. Programs ASTRA [23] and JANA [24] were used for additional data reduction and the following structure refinement. The refinement procedure is described in the next section. Small differences in the unit-cell values are not taken into account since they do not make a visible influence on the calculation results. As evident from Table 1, negative thermal expansion (NTE) is observed between 50 K and 135 K. The NTE phenomenon as well as unaveraged unit-cell values of $LuB_{12}$ at different temperatures will be discussed later in the separate work.

**Table 1**. Experimental details

| Chemical formula | $LuB_{12}$ | | | |
|---|---|---|---|---|
| Temperature (K) | 50 | 95 | 135 | 293 |
| Crystal diameter (mm) | 0.28 | 0.25 | | 0.28 |
| System, space group, Z | Cubic, $Fm\bar{3}m$, 4 | | | |
| $a_{cub}$ (Å) | 7.4581(1) | 7.4564(1) | 7.4561(1) | 7.4610(1) |
| $V$ (Å$^3$) | 414.84(1) | 414.56(1) | 414.51(1) | 415.33(1) |
| Radiation type | $MoK_\alpha$, $\lambda = 0.7107$ Å | | | |
| $\mu$ (mm$^{-1}$) | 2.360 | 2.362 | 2.362 | 2.358 |
| Data collection | | | | |
| Diffractometer | Huber-5042 | Xcalibur S | | CAD4 |
| $\theta_{max}$ (°) | 74.25 | 73.72 | 73.75 | 74.83 |
| Number of measured, independent, observed [$I > 3\sqrt{I}$] reflections; $R_{int}$ | 5544, 293, 293; 0.119 | 10825, 265, 265; 0.058 | 10823, 265, 265; 0.060 | 13565, 271, 271; 0.027 |
| Refinement method | LSQ based on $F^2$ | | | |
| Weighting scheme $w = 1/\sigma^2(I)+(kI)^2$, $k$ | 0.02 | 0.03 | 0.03 | 0.008 |
| Refined parameters | 7 | | | |
| R, wR, S | 0.014, 0.034, | 0.012, 0.028, | 0.012, 0.029, | 0.005, 0.009, |

|  | 1.29 | 0.96 | 0.98 | 0.97 |
|---|---|---|---|---|
| $\Delta\rho_{max}, \Delta\rho_{min}$ (e/Å$^3$) | 2.84, -1.1 | 0.47, -0.37 | 1.33, -0.56 | 0.49, -0.29 |

## 3. Calculations

### 3.1. Problem definition

Residual factor (R-factor) is a sound argument in discussions of the crystal symmetry, and in case of equal R-factors the more symmetrical model is recommended as containing fewer refined parameters. R-factor is usually lower if the structure of LuB$_{12}$ is refined in $Fm\bar{3}m$ than in a less symmetrical group. Moreover, atomic positions do not undergo visible shifts being refined in a less symmetrical model what means that the choice of $Fm\bar{3}m$ is predetermined. Complex structure factors $F_{calc} = |F_{calc}(\mathbf{H})|\exp[i\psi_{calc}(\mathbf{H})]$, which are calculated from the refined structure parameters, and observed $F_{obs} \cong |F_{obs}(\mathbf{H})| \exp[i\psi_{calc}(\mathbf{H})]$ are involved into the procedure of the difference Fourier synthesis of residual ED to reveal the additional structure details that are not taken into account yet. Residual electron density $\Delta g(\mathbf{r})$ at a point $\mathbf{r}$ of the unit cell is calculated from the following formula:

$$\Delta g(\mathbf{r}) = (1/V)\Sigma_{\mathbf{H}} \mid |F_{obs}(\mathbf{H})| - |F_{calc}(\mathbf{H})| \mid \exp[i\psi_{calc}(\mathbf{H})] \exp(-2\pi i\, \mathbf{Hr}), \qquad (1)$$

where $V$ is a unit-cell volume and the summation is over the reciprocal lattice points $\mathbf{H} = \mathbf{H}_{hkl} = h\mathbf{a}^* + k\mathbf{b}^* + l\mathbf{c}^*$.

Evidently, the formula (1) does not contain any information on the crystal symmetry and calculations of $\Delta g(\mathbf{r})$ can be performed at any point $\mathbf{r}$ of the unit cell. It is well known, however, that Fourier maps ideally reproduce prescribed symmetry. This happens because computational programs support ED calculations in a symmetry-independent part of the unit cell to speed up the summation procedure. The result is then expanded on the whole cell by the symmetry operators. And besides, equivalent-averaged modules of the structure factors can participate in calculations instead of initial ones. Three difference Fourier maps are calculated within the limits of $Fm\bar{3}m$ as an illustration (Fig. 2). The $x = 0$, $y = 0$, $z = 0$ planes are symmetry-equivalent in $Fm\bar{3}m$, so respective maps are identical. Central atom Lu(0,0,0) is surrounded with eight boron atoms. Other Lu atoms are in vertexes of a square whose sides are periods of the unit cell. Residual ED peaks stand out being oriented along the square diagonals at distances of ~0.5 Å from the Lu site, whose origin may be associated with disordering of cationic lattice.

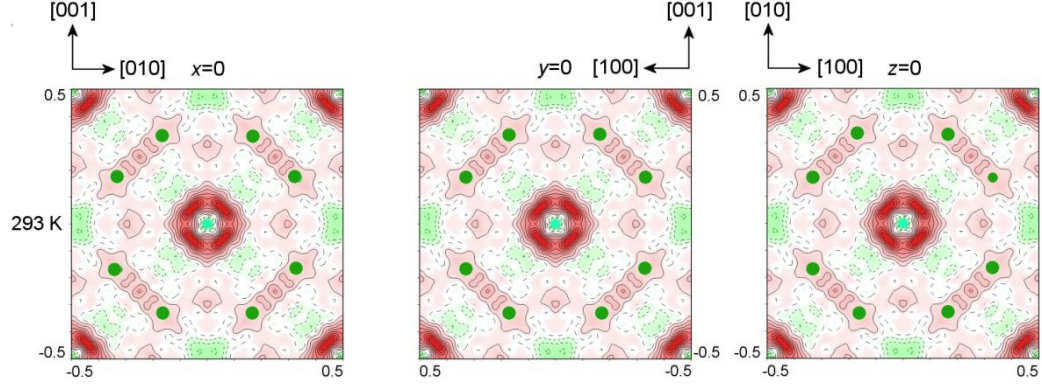

**Fig. 2**. Residual ED distribution in the $x = 0$, $y = 0$, $z = 0$ planes of LuB$_{12}$. Difference Fourier map is calculated in *Fm-3m* using data collected at 293 K for the sample #1. Contour intervals are 0.05 e/Å$^3$. Positive (pink online) and negative (green online) residual ED is highlighted. Central Lu(0,0,0) site (lime green circle online) is surrounded with eight boron sites (dark-green circles online); [-0.5, 0.5] intervals are periods of the crystal lattice.

Thus, on the one hand, $Fm\bar{3}m$ is the best symmetry group for the structure refinement. On the other hand, we cannot move forward on the explanation of the observed anisotropy of physical properties being limited by the symmetry restrictions coded in conventional software. As well as previous researchers, we can just affirm that cationic sublattice of LuB$_{12}$ is most likely disordered. A new idea is necessary to break the deadlock.

*3.2. Problem solution*

3.2.1. Accurate experimental data allow suggesting an approach to the residual ED calculations, which is mainly or only based on experimental data without leaning on the symmetry of the structure model. The idea is as follows. First of all, measured $|F_{obs}|$ are averaged in the $m\bar{3}m$ Laue class and the crystal structure of LuB$_{12}$ is refined in the $Fm\bar{3}m$ symmetry group according to a simple scheme. Independent atomic set in $Fm\bar{3}m$ consists of two atoms. The Lu atom is fixed in the 4*a* position as Lu$_{4a}$ (0,0,0). The boron atom is placed in B$_{48i}$ (~0.17,~0.33, 0). Both atoms are supplied with isotropic thermal parameters (atomic displacement parameters, ADPs in present-day nomenclature). So, refined structural parameters are two coordinates (*x, y*) of the boron atom and two isotropic ADPs of B and Lu atoms. Two more refined parameters are scale factor and isotropic extinction parameter. After the structure is refined, its atoms are considered in a less-symmetrical group. Atomic coordinates Lu and B are expanded to create a new independent set. Updated arrays of $|F_{obs}|$ и $|F_{calc}|$ have to be created for the future calculations since present $|F_{obs}|$ are averaged in the cubic $m\bar{3}m$ class and corresponding $|F_{calc}|$ inherit this feature. The new set of

|$F_{obs}$| is averaged in the less symmetrical Laue class whereas a good way to update |$F_{calc}$| is a 'refinement' of the updated structure model. The term 'refinement' is quoted since all atomic coordinates are fixed and all ADPs are kept isotropic and equal for all the atoms of one type. Only four refined parameters are thus left in the model, and no one of them can influence on the symmetry of the ED distribution. They are: isotropic ADPs of Lu and B, scale factor and isotropic extinction parameter.

3.2.2. After preparatory works are finished, one of two independent computational procedures may be performed. We made both as a rule to compare results. One of them is difference Fourier synthesis. Calculations are made in less symmetrical group within expanded part of the unit cell. The result is expected to be less symmetrical in case the crystal is not exactly cubic in reality. Unit-cell values are kept cubic since very small deviations of the lattice parameters from the cubic ones do not influence on the results.

We first applied such an approach to a crystal of $LuB_{12}$, whose structure was refined in *Fm$\bar{3}$m* but difference Fourier synthesis of residual ED was performed in the orthorhombic *Fmmm* group. Fourier synthesis in the less symmetrical group revealed actual absence of those fourfold axes, which were mandatory in *Fm$\bar{3}$m*. In this way we managed to explain an observed difference in magnetoresistance along [100] and [010] crystallographic axes, which are equivalent in a cubic crystal. Cooperative Jahn-Teller effect was considered as the most probable cause of the symmetry lowering [13]. In the present work, we went down from cubic to the least symmetrical group with the largest independent part of the unit cell to prevent an artificial symmetry overstating as far as possible. Systematical extinctions of the reflections do not give grounds to a renunciation of the *F*-centering whereas the renunciation of the inversion center does not lead to visible changes in the difference Fourier maps, so the difference Fourier synthesis was performed based on the data sets corresponding to the non-standard *F$\bar{1}$* group.

3.2.3. Another scheme is based on the maximal entropy method (MEM) [25]. The unit cell is divided into small volumes (voxels) and a MEM-reconstructed ED value of $g_{MEM}$ is assigned to each of them. The method operates directly with structure factors $F_{obs}$ and $F_{calc}$ but not with atomic coordinates or ADPs. MEM does not even demand chemical composition of the crystal being guided by general number of electrons in the unit cell. As well as above, the calculations are limited by independent part of the unit cell depending on symmetry of the structure model. Because of that, the MEM calculations were performed for *F$\bar{1}$* by analogy with difference Fourier synthesis. The MEM calculations were made by Dysnomia program (see [26] for the MEM formalism and computational details). The program VESTA [27] was used for visualization of the MEM results.

**4. Results and discussion**

*4.1. Asymmetry of residual ED distribution*

Selected results of the difference Fourier synthesis of the residual ED in LuB$_{12}$ are presented in Fig. 3. The calculations are made in the $F\bar{1}$ group using data sets collected at four temperatures. Three maps in each horizontal row are drawn in the $x = 0$, $y = 0$, $z = 0$ planes of the crystal lattice.

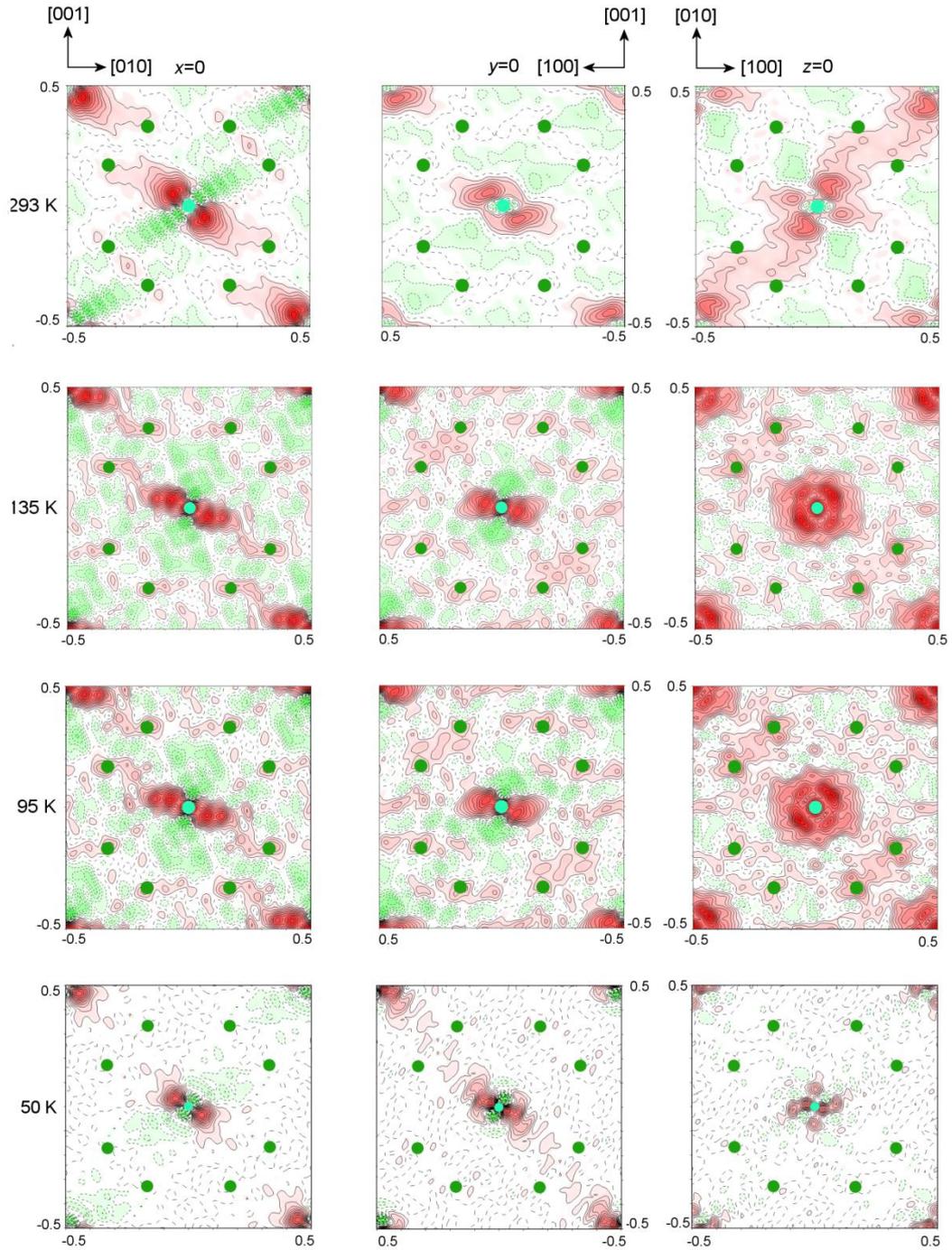

**Fig. 3**. Residual ED distribution in the $x = 0$, $y = 0$, $z = 0$ planes of the LuB$_{12}$. Difference Fourier synthesis is done in $F\bar{1}$ using data collected at four temperatures. Contour intervals are 0.2 e/Å$^3$ (295, 135, 95 K) and 1 e/Å$^3$ (50 K). Positive (pink online) and negative (light-green online) residual ED is highlighted. Central Lu(0,0,0) site (lime green circle online) is surrounded with eight boron sites (dark green circles online); [-0.5, 0.5] intervals are periods of the crystal lattice.

As follows from the Fig. 3, the maps do not contain some of twofold axes, which are mandatory for each orthorhombic crystal, what means that the symmetry of the residual ED distribution is lower than even orthorhombic. At the same time, maps in the columns reveal an evident resemblance although they are obtained for two crystals based on data collected at different temperatures on different diffractometers. Residual ED near Lu(0,0,0) in the $x = 0$ plane is elongated being directed between the space diagonal $[01\bar{1}]$ and the edge [010] of the unit cell as if the central lutetium atom had a preferred contact with two of eight boron atoms. Similarly, elongated ED in the $y = 0$ plane is oriented between [101] and [100] towards two other boron atoms. In the $z = 0$ plane, residual ED peaks are bound by a likeness of fourfold axis what recurs to a thought of a tetragonal distortion of the crystal. Let's remember that distortions of the lattice parameters $a \cong b > c$ are of tetragonal type although the difference between the parameters is less than 0.01 Å. The same maps show signs of a trigonal distortion. In all rows in Fig. 3 except the bottom one, stronger two of four residual peaks in the plane $z = 0$ are oriented roughly along $[110] \equiv [\bar{1}\bar{1}0]$. Thus, residual ED is preferably located near the side diagonals $[\bar{1}\bar{1}0]$, $[01\bar{1}]$, [101], which are bound by the perpendicular space diagonal $[\bar{1}11]$ as if it were the threefold axis. Contour interval in the difference Fourier maps is increased by a factor of five in the bottom row as compared to upper rows. It must be taken into account to form a true judgment on temperature dependence of the residual ED peaks. Although the maps in the bottom row look pale, the peaks grow with temperature decreasing.

*4.2. MEM results*

We realize that LuB$_{12}$ is generally considered as an exemplary cubic crystal. Additional calculations have therefore been made to check previous results. Independent information has been obtained by MEM based on the same diffraction data sets. The program VESTA assures the 2D visualization of the results by cutting out a thin layer in the 3D array of $\rho_{MEM}$ (Fig. 4). MEM reconstructs a 'normal' but not difference ED so that light boron atoms are clearly seen in Fig. 4 just as heavy lutetium atoms. ED in the layer of any given thickness is automatically divided into ten levels from zero to $\rho_{max}$, each of them is assigned to a definite color from dark-blue to red. The values of $\rho_{MEM}$ in Fig. 4 are cut at the level $g_{max} = 0.1\%$ of the peak of $g_{MEM}(Lu)$ to show fine ED gradations in the thin layer.

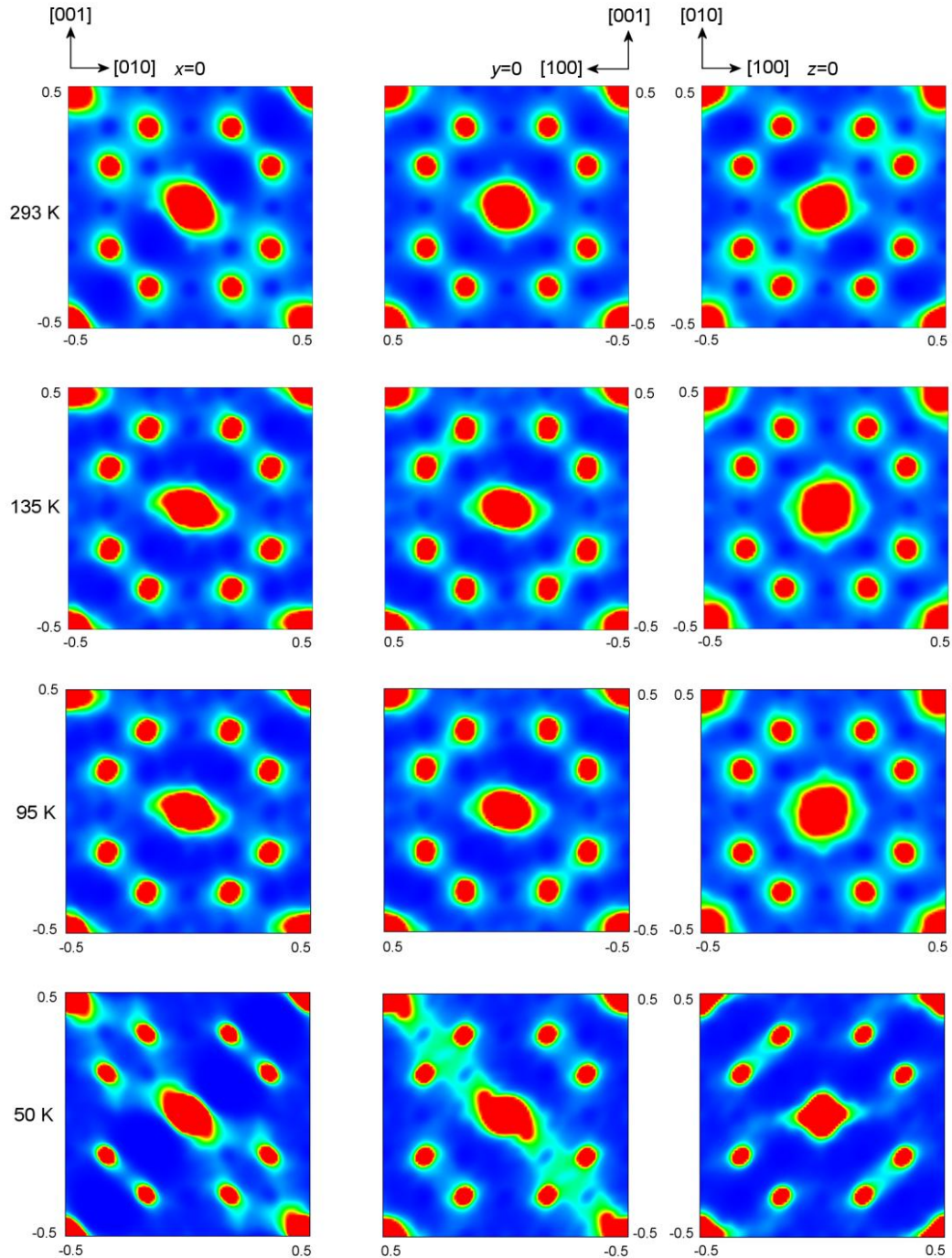

**Fig. 4**. MEM maps are calculated from the data sets collected at temperatures 293, 135, 95 and 50 K. Three columns from left to right present thin slices of ED distribution in three planes of the crystal lattice. Central Lu is surrounded by eight boron atoms; [-0.5, 0.5] intervals are periods of the crystal lattice.

First of all, it should be mentioned that the symmetry of 'normal' ED (Fig. 4) is not cubic at all temperatures studied. Moreover, corresponding maps in Figs. 3&4 reveal an evident similarity in character of the ED distribution. The middle map in the bottom row of Fig. 4 contains the most evident indications of static shifts of some cations from the $Lu_{4a}$ position in the [101] direction that allows to make a reference to the Yb$^{3+}$ – Yb$^{3+}$ pairs bounded in a certain direction in the crystal of YbB$_{12}$ [14].

*4.3. 'Structure – properties' relationship*

As for LuB$_{12}$, a diagonal stripe of high ED (Fig. 4, 50 K, $y = 0$) looks like conducting channel and it becomes of interest, therefore, to compare results of MEM and difference Fourier synthesis with peculiarities of the charge transport in the crystals. The resistivity ρ(T) curves for LuB$_{12}$ single crystal are shown (Fig.5a) in the absence of an external magnetic field (curve 1) and in steady field of 8 Tesla directed along the crystal axes, **B** ∥ [1$\bar{1}$0], [0$\bar{1}$0] and [110] (curves 2, 3 and 4, correspondingly). A significant anisotropy (up to 8%) of the transverse magnetoresistance at B = 8 T is observed below T* ≈ 60 K for **B** ∥ [1$\bar{1}$0] and **B** ∥ [110] (see curves 2 and 4 in Fig.5b for comparison). Fig. 5c shows an angular dependence of magnetoresistance measured at fixed temperature T = 4.2 K in the experiment with the sample rotating around its current axis **I** ∥ [001], which is perpendicular to **B**. Both **I** and **B** vectors are fixed whereas the normal **n** to the preselected crystal face (*hk*0) rotates around [001], so **B** becomes parallel to one or other [*hk*0] axis in the crystal while an angle φ between **n** and **B** changes from 0 to 360°. Despite the fact that [1$\bar{1}$0] and [110] axes are symmetry-equivalent in a cubic crystal, the corresponding ρ-values are maximal near φ = 0 & 180° and minimal near φ = 90 & 270° (see Fig. 5c).

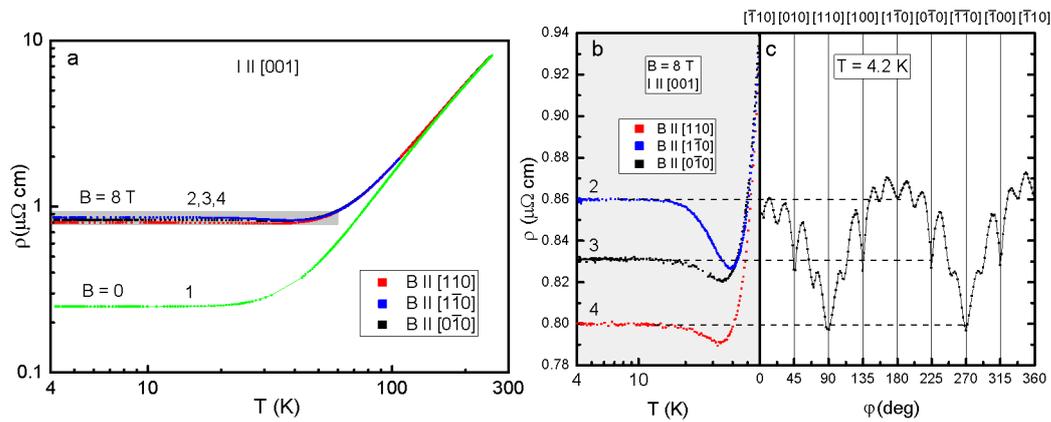

**Fig. 5.** (a) Temperature dependences of resistivity for LuB$_{12}$ single crystal (**I** ∥ [001]) in the absence of an external magnetic field (curve 1) and in steady field of 8 T directed along the crystal axes, **B** ∥ [1$\bar{1}$0], [0$\bar{1}$0] and [110] (curves 2, 3 and 4, correspondingly). (b) Scaled-up fragment of (a); (c) angular dependence of the transverse magnetoresistance ρ measured with B = 8 T at low temperature 4.2 K.

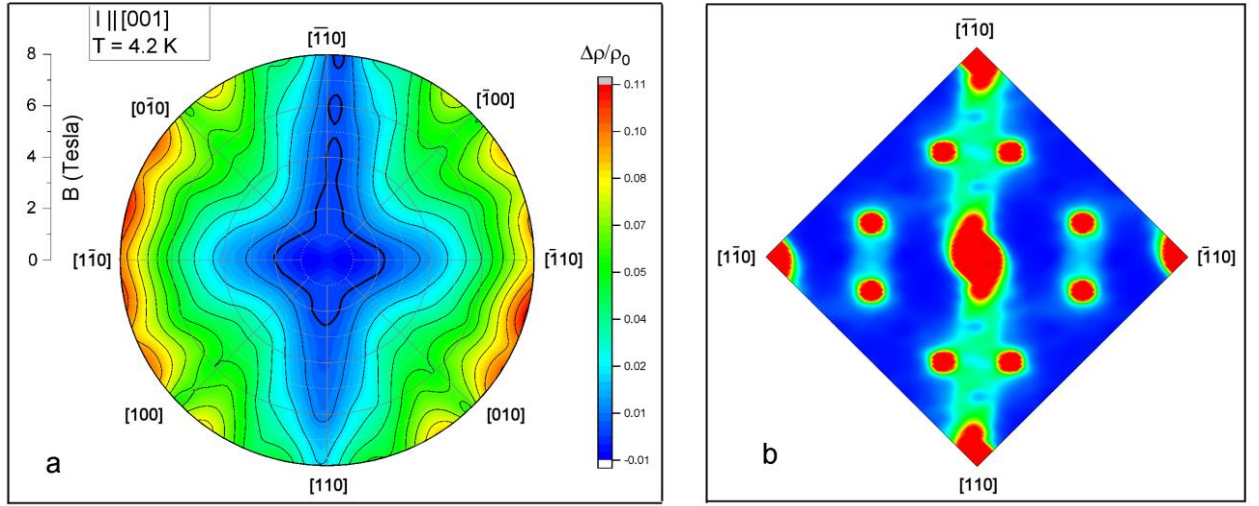

**Fig. 6**. (a) Anisotropy of magnetoresistance of LuB$_{12}$ in polar coordinates: $\Delta\rho/\rho_0 = [\rho(\varphi, B) - \rho(\varphi_0, B)]/\rho(\varphi_0, B)$, $\varphi_0 = 270°$ corresponding to **B** ∥ [$\bar{1}\bar{1}0$]; (b) anisotropic ED distribution in a thin layer of ED reconstructed by MEM.

The anisotropy of magnetoresistance ρ can be figured in polar coordinates as presented in Fig. 6a, where the magnetic field induction B changes along radii from 0 to 8 T, the angle φ changes circle-wise between various crystallographic directions and color shows the magnetoresistance amplitude. It is easy to discern from the plot that minimal values of magnetoresistance at any B are distributed mainly in the [110] direction and Δρ/ρ(φ$_0$) increases essentially at higher B values in the orientation of magnetic field **B** ∥ [1$\bar{1}$0]. The 'zero-Δρ/ρ$_0$' stripe in Fig. 6a is oriented in an exact correspondence, including a small deviation from the [110] direction, with the ED channel (charge stripe) in Fig. 6b, where the middle MEM map from the bottom row in Fig. 4 is presented being 45° turned to guide the reader's eye. The X-ray data were collected at higher temperature 50 K. Moreover, the data were collected in the absence of an external magnetic field (B = 0) but the expected anisotropy of conducting properties follows from Fig. 6b and shows itself as the visible anisotropy of transverse magnetoresistance when appropriate conditions are satisfied.

**Conclusion**

Accurate structure analysis of LuB$_{12}$ single crystals has been performed at temperatures 293, 135, 95, 50 K based on the X-ray data of high quality. A peculiar strategy is developed, which makes it possible to detect probable asymmetry of the ED distribution within the unit cell of dodecaborides. New approach consists in the difference Fourier synthesis of the residual ED as well as in the reconstruction of the ED distribution using MEM. The calculations are organized so as if the symmetry-equivalent parts of the unit cell were symmetry-independent.

The asymmetrical ED distribution near the Lu sites has been revealed using both these techniques. Residual ED is distributed mainly along the side-diagonals linked by a space diagonal of the cubic cell (one of threefold axes of $Fm\bar{3}m$). Residual peaks are distributed so that an influence of a fourfold axis is visible (one of fourfold axes of $Fm\bar{3}m$). The ED peaks become stronger with temperature decreasing and form a filamentary structure of conduction channels – unbroken charge stripes almost along the [110] axis.

These observations are in accordance with the cubic symmetry distortions of the $LuB_{12}$ crystals observed before. It has been shown that the asymmetrical ED distribution correlates very accurately with anisotropy of magnetoresistance. The same conduction channel is observed from the X-ray and charge-transport data, whose orientation in the unit cell is determined right up to a small deviation from the [110] direction.

**Acknowledgements**

This work was supported by the Federal Agency of Scientific Organizations [Agreement No 007-GZ/Ch3363/26] in part of X-ray diagnostics, by the Russian Foundation for Basic Research [grant no. 16-02-00171] in part of single-crystal structure analysis and by the Program of Fundamental Research of the Presidium of the Russian Academy of Sciences 'Fundamental Problems of High-Temperature Superconductivity' in part of resistance measurements and analysis. The X-ray data were collected using the equipment of the Shared Research Center FSRC 'Crystallography and Photonics' RAS whose work was partially supported by the Russian Ministry of Education and Science.